\documentstyle[graphicx,amsmath]{elsart}

\newcommand{\vecbm}[1]{\mbox{\boldmath$#1$}}

\newcommand{\lra} {$\leftrightarrow$}
\newcommand{\vecb}[1]{\mbox{\bf#1}} 
\newcommand{\lora}{{\boldmath$\longrightarrow$}}
\begin{document}
\title{Non-extensive Hamiltonian systems follow Boltzmann's principle
  not Tsallis statistics. -- Phase Transitions, Second Law of
  Thermodynamics} \author{D.H.E. Gross} \address{Hahn-Meitner-Institut
  Berlin, Bereich Theoretische Physik,Glienickerstr.100\\ 14109
  Berlin, Germany and Freie Universit{\"a}t Berlin, Fachbereich
  Physik; \today}
{\scriptsize Abstract: Boltzmann's principle $S(E,N,V)=k\ln W(E,N,V)$
  relates the entropy to the geometric area $e^{S(E,N,V)}$ of the
  manifold of constant energy in the N-body phase space. From the
  principle all thermodynamics and especially all phenomena of phase
  transitions and critical phenomena can be deduced. The topology of
  the curvature matrix $C(E,N)$ (Hessian) of $S(E,N)$ determines
  regions of pure phases, regions of phase separation, and
  (multi-)critical points and lines. Thus, $C(E,N)$ describes all kind
  of phase-transitions with all their flavor. No assumptions of
  extensivity, concavity of $S(E)$, additivity have to be invoked.
  Thus Boltzmann´s principle and not Tsallis statistics describes the
  equilibrium properties as well the approach to equilibrium of
  extensive and non-extensive Hamiltonian systems. No
  thermodynamic limit must be invoked}\\ PACS numbers:
05.20.Gg,05.70Ln
\section{Introduction}
There are many attempts to derive Statistical Mechanics from first
principles. The earliest are by Boltzmann~\cite{boltzmann1877},
Gibbs~\cite{gibbs36}, and Einstein~\cite{einstein04}. The two central
issues of Statistical Mechanics according to the deep and illuminating
article by Lebowitz~\cite{lebowitz99a} are to be explained: How
irreversibility (the Second Law of Thermodynamics) arises from fully
reversible microscopic dynamics, and the other astonishing phenomenon
of Statistical Mechanics: the appearance of phase transitions.
Moreover, conventional statistical mechanics works in the
thermodynamic limit. I.e. all non-extensive, "Small" systems having
linear dimensions comparable to the range of the forces are excluded.

\section{Minimum-bias deduction of Statistical Mechanics}
Thermodynamics presents an economic but reduced description of a
$N$-body system with a typical size of $N\sim 10^{23}$ particles in
terms of a very few ($M\sim 3-8$) ``macroscopic'' degrees of freedom
($dof$s). In order to address also "Small" systems I will allow for
much smaller systems of some $100$ particles like nucleons in a
nucleus. However, I assume that always $6N\gg M$.  The believe that
phase transitions and the Second Law can exist only in the
thermodynamic limit turns out to be false.

Evidently, determining only $M$ $dof$s leaves the overwhelming number
$6N-M$ $dof$s undetermined. {\em All} N-body systems with the same
macroscopic constraints are {\em simultaneously} described by
Thermodynamics. These systems define an {\em ensemble}
$\cal{M}$~\footnote{In this paper I denote ensembles or manifold in
  phase space by calligraphic letters like ${\cal{M}}$.} of points in
the $N$-body phase space. Thermodynamics can only describe the {\em
  average} behavior of this whole group of systems. I.e.\ it is a {\em
  statistical} or {\em probabilistic} theory. Macroscopic quantities
are averages over ${\cal{M}}$.

The dynamics of the (eventually interacting) $N$-body system is ruled
by its Hamiltonian $\hat{H}_N$. Let us in the following assume that
our system is trapped in an inert rectangular box of volume $V$ and
there is no further conservation law than the total energy. The motion
in time of all points of the ensemble follows trajectories in $N$-body
phase space $\{q_i(t),p_i(t)\}|_{i=1}^N$ which will never leave the
($6N-1$)-dimensional shell (or manifold) $\cal{E}$ of constant energy
$E$ in phase space.  We call this manifold the {\em micro-canonical}
ensemble.

An important information which contains the whole equilibrium
Statistical Mechanics including all phase transition phenomena is the
area $W(E,N) =:e^S$ of this manifold $\cal{E}$. Boltzmann has shown
that $S(E,N)$ is the {\em entropy} of our system. Thus the entropy
and with it equilibrium thermodynamics has a {\em geometric}
interpretation.

Einstein called Boltzmann's definition of entropy as e.g.\ written on
his famous epitaph
\begin{equation}
\fbox{\fbox{\vecbm{$S=k$\cdot$lnW$}}}\label{boltzmentr1}\end{equation}
{\em Boltzmann's principle}~\cite{einstein05d} from which Boltzmann was
able to deduce thermodynamics. Precisely $W$ is the number of
micro-states of the $N$-body system at given energy $E$ in the
spatial volume $V$ and further-on I put Boltzmann's constant $k=1$:
\begin{eqnarray}
W(E,N,V)&=& tr[\epsilon_0\delta(E-\hat H_N)]\label{partitsum}\\
tr[\delta(E-\hat H_N)]&=&\int_{\{q_\alpha\in V\}}{\frac{1}{N!}
\left(\frac{d^3q_\alpha\;d^3p_\alpha}
{(2\pi\hbar)^3}\right)^N\delta(E-\hat H_N)},\label{phasespintegr}
\end{eqnarray} 
$\epsilon_0$ is a suitable energy constant to make $W$ dimensionless,
the $N$ positions $q_\alpha$ are restricted to the volume $V$, whereas
the momenta $p_\alpha$ are unrestricted. In what follows, I remain on
the level of classical mechanics. The only reminders of the underlying
quantum mechanics are the measure of the phase space in units of
$2\pi\hbar$ and the factor $1/N!$ which respects the
indistinguishability of the particles (Gibbs paradoxon). With this
definition, eq.\ref{boltzmentr1}, {\em the entropy $S(E,N,V)$ is an
  everywhere multiple differentiable, one-valued function of its
  arguments.} This is certainly not the least important difference to
the conventional canonical definition.

In contrast to Boltzmann~\cite{boltzmann1877} who used
the principle only for dilute gases and to
Schr\"odinger~\cite{schroedinger44}, who thought equation
(\ref{boltzmentr1}) is useless otherwise, I take the principle as
{\em the fundamental, generic definition of entropy}. In a recent
book~\cite{gross174} cf.\ also~\cite{gross173,gross175} I
demonstrated that this definition of thermo-statistics works well
especially also at higher densities and at phase transitions without
invoking the thermodynamic limit. This is important: Elliot
Lieb~\cite{lieb97} considers the additivity of $S(E)$ and
Lebowitz~\cite{lebowitz99a,lebowitz99} the thermodynamic limit as
essential for the deduction of thermo-statistics. However, neither is
demanded if one starts from Boltzmann's principle.

{\em This is all that Statistical Mechanics demands}, no further
assumption must be invoked. Neither does one need extensivity, nor
additivity, nor concavity of $S(E)$ c.f.~\cite{lavanda90}.{\em
  Boltzmann's principle eq.(\ref{boltzmentr1}) is the only axiomatic
  assumption necessary for thermo-statistics.}
\section{The micro-canonical ensemble is the fundament ensemble}
\label{fundam} During the dynamical evolution of a many-body system
interacting by short-range forces the internal energy is conserved.
Only perturbations by an external ``container'' can change the energy.
I.e.\ the fluctuations of the energy are $\frac{\Delta E}{E}\propto
V^{-1/3}$.  If, however, the diameter of the system is of the order of
the range of the force, i.e.\ the system is ``Small'', non-extensive,
details of the coupling to the container cannot be ignored. The
canonical ensemble does not care about these details, assumes the
system is homogeneous, averages over a Boltzmann-Gibbs (exponential)
distribution $P_E\{q_\alpha,p_\alpha\}=\frac{1}{Z(\beta)}e^{-\beta
  \hat H\{q_\alpha,p_\alpha\}}$ of energy and fixes only the mean
value of the energy by the temperature $1/\beta$. In order to agree
with the micro, $e^{-\beta E}W(E)$ must be sharp in $E$ i.e.
self-averaging, what is usually not the case in non-extensive systems.
Then one must work in the micro, the only orthode ensemble.

In this conference it is suggested to describe the equilibrium of
non-extensive systems by Tsallis
q-entropy~\cite{tsallis88,martinez00a,martinez00,abe01}:
\begin{equation}
S_q=k\frac{1-\sum_{i=1}^W{P_i^q}}{q-1}.
\end{equation}
For a closed {\em Hamiltonian} system at energy $E$, the $P_i$ are the
probabilities for each of the $W(E)$ microscopic configurations
(quantum states).  Following Toral~\cite{toral01} this has of course the
following consequences: After maximizing $S_q(E)$ under variation of
$P_i$ with the constraint of $\sum{P_i}=1$ one obtains the equal
probability distribution characterized by Boltzmann's entropy
$W(E)=e^{S(E)}$:
\begin{equation}
P_i(\epsilon)=\left\{\begin{array}{ll} e^{-S(E)}&,\epsilon_i=E\\
                     0&,otherwise
    \end{array}\right.,\hspace{1cm}S_q=k\frac{1-e^{(1-q)S(E)}}{q-1}.
\end{equation}
Moreover, following Abe and Toral~\cite{abe01,toral01} the original
definitions of the microcanonical temperature and pressure,
eq.~(\ref{TP}) below, through Boltzmann's entropy $S(E,N,V)$,
eq.(\ref{boltzmentr1}), are the only way within Tsallis statistics to
define the equilibrium of two systems in weak contact and to fulfill
the zeroth law under energy- and volume exchange:
\begin{eqnarray}
S(E,N,V)&=&\ln W(E,N,V)\\ 
T_{phys}=\left(\frac{\partial S}{\partial E}\right)^{-1}&\hspace{1cm}&
P_{phys}=\frac{\partial S/\partial V}{\partial S/\partial E}\label{TP}
\end{eqnarray}
I.e. the physical quantity relevant for {\em equilibrium} of {\em
  Hamiltonian} systems, extensive or not, is the original Boltzmann
entropy $S(E)=\ln[W(E)]$, eq.(\ref{boltzmentr1}), whatever the
non-extensivity index $q$.  Therefore, for closed {\em Hamiltonian}
many-body systems {\em at statistical equilibrium}, extensive or
not,{\em the thermo-statistical behavior is entirely controlled by
  Boltzmann's principle and the microcanonical ensemble as discussed
  in this paper.} Tsallis statistics seems to apply to non-equilibrium
situations like turbulence etc.
\section{Phase transitions within Boltzmann's principle}
At phase-separation the system becomes inhomogeneous and splits into
different regions with different structure. This is the main generic
effect of phase transitions of first order.  Evidently, phase
transitions are foreign to the (grand-) canonical theory with
homogeneous density distributions.  Therefore, in the conventional
Yang-Lee theory phase transitions~\cite{yang52} are indicated by the
zeros of the grand-canonical partition sum where the grand-canonical
formalism breaks down (Yang--Lee singularities).

Within Boltzmann´s principle phase-transitions are generically
classified in terms of the topology of curvatures of $S(E,N)$ i.e. by
its Hessian and its eigenvalues $\lambda_{1,2}$. This works also for
``Small'', non-extensive systems, details in~\cite{gross174}:
\begin{equation}
\det(E,N)= \left\|\begin{array}{cc}
\frac{\partial^2 S}{\partial E^2}& \frac{\partial^2 S}{\partial N\partial E}\\
\frac{\partial^2 S}{\partial E\partial N}& \frac{\partial^2 S}{\partial N^2}
\end{array}\right\|= \left\|\begin{array}{cc} S_{EE}&S_{EN}\\
S_{NE}&S_{NN}
\end{array}\right\|=\lambda_1\lambda_2,\;\lambda_1\ge\lambda_2,
 \label{curvdet}
\end{equation}
\begin{itemize}
\item A {\bf single stable} phase by $\lambda_1<0$.  Here $S(E,N)$ is
  locally concave (downwards bended) in both directions and if the
  $T,\mu$-anologue to eqs.(\ref{TP}) have a single solution
  $E_s,N_s$, then there is a one to one mapping of the grand-canonical
  \lra the micro-ensemble.
\item A {\bf transition of first order} with phase separation and
  surface tension is indicated by $\lambda_1>0$.  $S(E,N)$ has a
  convex intruder (upwards bended) in the direction $\vecb{v}_1$ of
  the largest curvature (order parameter). The depth of the intruder
  is a measure of the inter-phase surface tension. Then the
  $T,\mu$-analogue to eqs.(\ref{TP}) have multiple solutions, at
  least three.  The whole convex area of \{E,N\} is mapped into a
  single point ($T,\mu$) in the grand-canonical ensemble
  (non-locality)\label{convex}. I.e.\ here {\em both ensembles are not
    equivalent and the (grand-)canonical ensemble is non-local in the
    order parameter and violates basic conservation
    laws}~\cite{gross174,gross173,gross175,gross180}. The region in
  \{$E,N$\} of separation of different phases, $\lambda_1>0$, is
  bounded by lines with $\lambda_1(E,N)=0$.
\item On this boundary is the end-point of the transition of first
  order, here the three solutions of eqs.\ref{TP} move into one
  another in the direction $\vecb{v}_1$, into the critical end-point
  of the first order transition ($\vecb{v}_{1,2}$ are the eigenvectors
  of the Hessian). Here, we have a {\bf continuous (``second order'')}
  transition with vanishing surface tension, where two neighboring
  phases become indistinguishable, here $\lambda_1(E,N)=0$ and
  $\vecb{v}_1\cdot\vecbm{\nabla}\lambda_1=0$.  These are the {\em
    catastrophes} of the Laplace transform $E\to T$.  Furthermore,
  there may be also whole lines of second-order transitions like in
  the antiferro-magnetic Ising model c.f.\cite{gross174}.
\item Finally, a {\bf multi-critical point} where more than two phases
  become indistinguishable is at the branching of several lines in the
  \{$E,N$\}-phase-diagram with $\lambda_1=0$, {\boldmath$\vecbm{\nabla}$}
   \mbox{$\lambda_1$}{\boldmath$=0$}.
\end{itemize}
\section{Fractal distributions in phase space, approach to equilibrium, 
  Second Law}\label{fractSL} 
\begin{figure}[b]
\begin{minipage}[b]{6cm}
\begin{center}$V_a$\hspace{2cm}$V_b$\end{center}
\vspace*{0.2cm}
\includegraphics*[bb = 0 0 404 404, angle=-0, width=5.7cm,
clip=true]{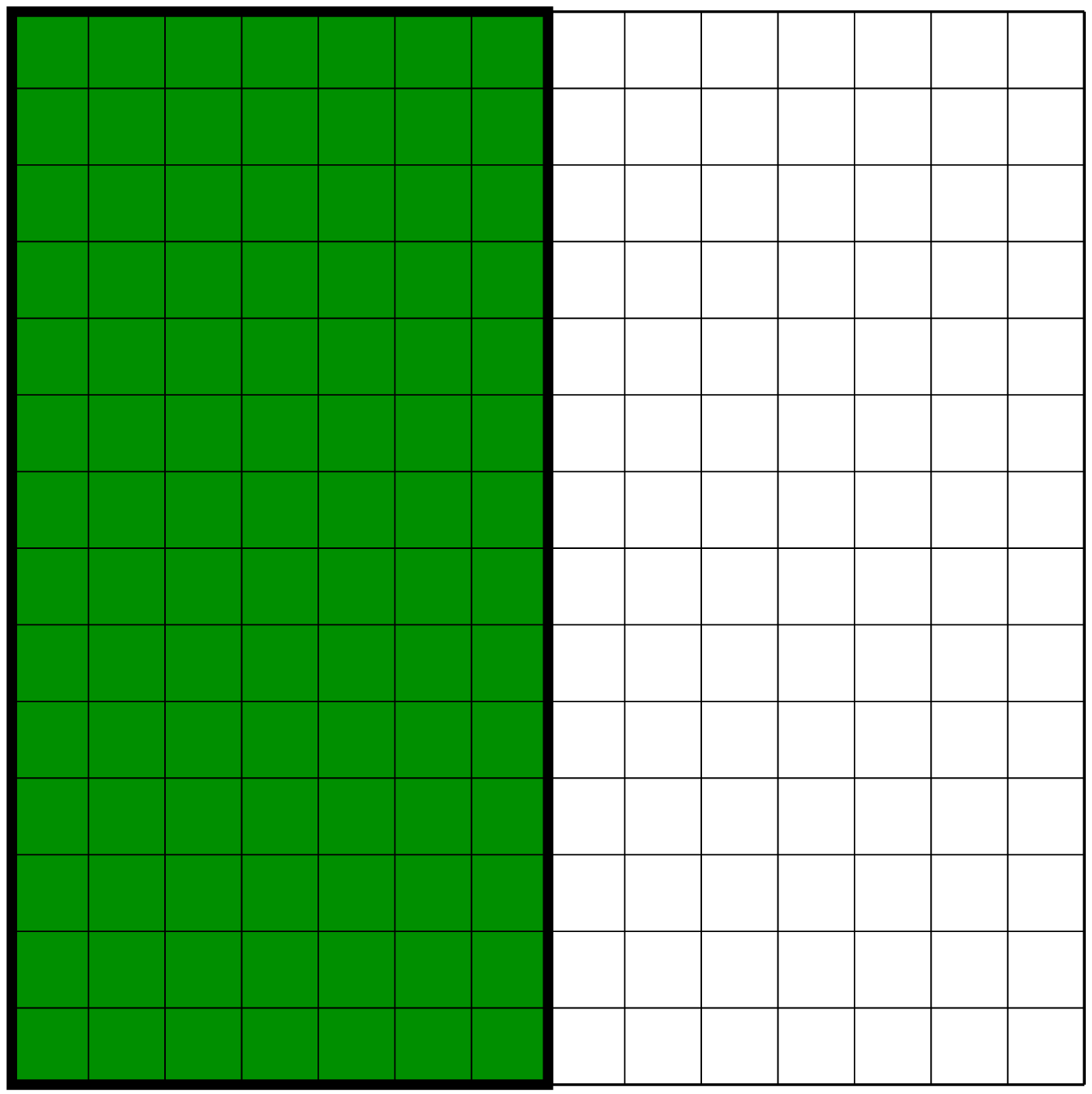}\begin{center}$t<t_0$\end{center}
\end{minipage}\lora\begin{minipage}[b]{6cm}
\begin{center}$V_a+V_b$\end{center}\vspace{-0.8cm}
\includegraphics*[bb = 0 0 490 481, angle=-0, width=6.9cm,
clip=true]{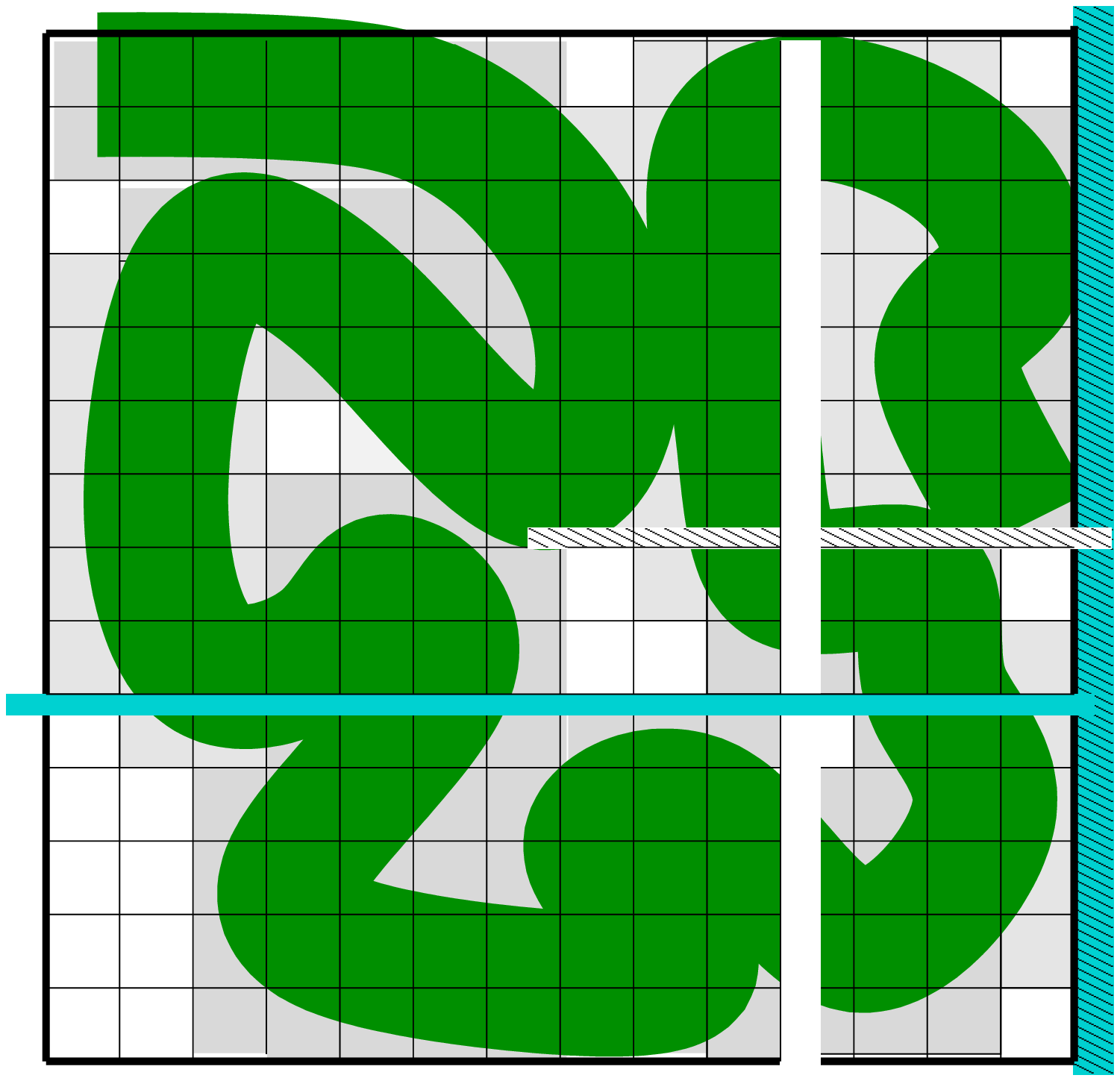}\begin{center}$t>t_0$\end{center}
\end{minipage}\\
\caption{The compact set ${\cal{M}}(t_0)$, left side, develops into an
  increasingly folded ``spaghetti''-like distribution
${\cal{M}}(t,t_0)$ in the phase-space with rising time $t$.  The
right figure shows only the early form of the distribution. At much
later times it will become more and more fractal and finally dense in
the new phase space.  The grid illustrates the boxes of the
box-counting method.  All boxes which overlap with ${\cal{M}}(t,t_0)$
contribute to the box-counting volume $\mbox{vol}_{box,\delta}$ and
are shaded gray.  Their number is $N_\delta$\label{secondL}}
\end{figure}

Let us examine the following Gedanken
experiment: Suppose the probability to find our system at points
$\{q_t,p_t\}_1^N$ in phase space is uniformly distributed at time
$t_0$ over the sub-manifold ${\cal{M}}(t_0,t_0)\equiv{\cal{E}}(N,V_1)$
of the $N$-body phase space at energy $E$ and spatial volume $V_1$. At
time $t>t_0$ we allow the system to spread over the larger volume
$V_2>V_1$ without changing its energy.  If the system is {\em
  dynamically mixing}, the majority of trajectories $\{q_t,p_t\}_1^N$
in phase space starting from points $\{q_0,p_0\}_1^N$ with $q_0\in
V_1$ at $t_0$ will now spread over the larger volume $V_2$.  Of course
the Liouvillean measure of the distribution ${\cal{M}}(t,t_0)$ in
phase space at $t>t_0$ remains the same ($=tr[{\cal{E}}(N,V_1)]$).
But as already argued by Gibbs the distribution ${\cal{M}}(t,t_0)$
will be filamented like ink in water and will approach any point of
${\cal{E}}(N,V_2)$ arbitrarily close.
$\lim_{t\to\infty}{\cal{M}}(t,t_0)$ becomes dense in the new, larger
${\cal{E}}(N,V_2)$.  The closure $\overline{{\cal{M}}(t=\infty,t_0)}$
becomes equal to ${\cal{E}}(N,V_2)$.  This is clearly expressed by
Lebowitz~\cite{lebowitz99a,lebowitz99}, and illustrated by the figure
\ref{secondL}. Thermodynamics cannot distinguish ${\cal{M}}$ from
$\overline{{\cal{M}}}$. {\em Thus the closed manifolds
  $\overline{{\cal{M}}}$ are the real objects of thermodynamics}.

To calculate $\overline{{\cal{M}}}$ we introduce the box-counting
``measure'' $M_\delta\propto N_\delta \delta^{6N-1}$ for the
distribution in phase-space as indicated in fig.\ref{secondL}. We
cover ${\cal{M}}$ with a rectangular grid with spacing $\delta$ and
and count the number $N_\delta$ of boxes which overlap with
${\cal{M}}$ with the convention taking $\delta\to 0$ after taking
averages, see \cite{gross180,gross182}. For finite times because of
Liouville's theorem we have
\begin{equation}
\underbar{$\lim$}_{\delta\to 0}M_\delta(t,t_0)=W(E,N,t_0,t_0)=W(E,N,V_1)
\end{equation}
At $t\to\infty$ the two limits $\delta\to 0,t\to\infty$ do in general
not commute and as assumed by Gibbs the manifold
${\cal{M}}(t\to\infty,t_0)$ becomes dense in the new micro-canonical
manifold ${\cal{E}}(V_2)$.  Then
\begin{equation}
\underbar{$\lim$}_{\delta\to 0}\lim_{t\to\infty}M_\delta(t,t_0)=W(E,N,V_2)\ge
W(E,N,V_1).
\end{equation}
{\em This is the Second Law of Thermodynamics.} For a more detailed mathematical
discussion c.f. \cite{gross180,gross182}.
\section{Conclusion}\label{conclus}
Macroscopic measurements $\hat{M}$ determine only a very few of all
$6N$ $dof$s.  Any macroscopic theory like thermodynamics deals with
the {\em volumes} $W=e^S$ of the corresponding closed sub-mani\-folds
$\overline{\cal{M}}$ in the $6N$-dim.\ phase space not with single
points.  The averaging over ensembles or finite sub-manifolds in phase
space becomes especially important for the micro-canonical ensemble of
a {\em non-extensive} or any other non-self-averaging system. Entropy
$S(E,N,V)$ is the natural measure of the geometric size of the
ensemble. The topology of its curvature indicates all phenomena of
phase transitions independently of whether the system is extensive or
non-extensive.

Several miss-interpretation of Statistical Mechanics are pointed out:
The existence of phase transitions and critical phenomena are {\em
  not} linked to the thermodynamic limit. They exist clearly and
sharply in ``Small'', non-extensive systems as well. Non-extensive
{\em Hamiltonian} systems do {\em not} demand a new entropy formalism
like that by Tsallis~\cite{tsallis88,vives01}. Boltzmann's principle,
the microcanonical ensemble, covers all equilibrium properties as well
as the approach towards equilibrium i.e. the Second Law.

By our derivation of micro-canonical Statistical Mechanics for finite,
non-extensive systems various non-trivial limiting processes are
avoided.  Neither does one invoke the thermodynamic limit of a
homogeneous system with infinitely many particles nor does one rely on
the ergodic hypothesis of the equivalence of (very long) time averages
and ensemble averages. As Bricmont~\cite{bricmont00} remarked
Boltzmann's principle is the most conservative way to Thermodynamics
but more than that it is the most straight one also.  {\em The single
  axiomatic assumption of Boltzmann's principle, which has a straight
  geometric interpretation, leads to the full spectrum of equilibrium
  thermodynamics including all kinds of phase transitions and
  including the Second Law of Thermodynamics.}

\end{document}